\documentstyle[twocolumn,prl,aps,epsf]{revtex}

\begin{document}\twocolumn[\hsize\textwidth\columnwidth\hsize\csname @twocolumnfalse\endcsname
\title{Auger transition from orbitally degenerate systems: Effects of
screening and multielectron excitations}

\author{D. D. Sarma$^{*}$ and Priya Mahadevan$^{\xi}$}
\address{Solid State and Structural Chemistry Unit, Indian Institute of 
Science, Bangalore 560012, India}
\date{\today}
\maketitle
\begin{abstract}
We calculate Auger spectra given by the two-hole Green's function
from orbitally degenerate Hubbard-like models as a function of 
correlation strength and band filling. The resulting spectra are
{\it qualitatively} different from those obtained from fully-filled singly
degenerate models due to the presence of screening dynamics and 
multielectron excitations. Application to a real system shows
remarkable agreement with experimental results leading to
reinterpretation of spectral features.
\end{abstract}

]

More than seventy years after its discovery \cite{auger}, Auger processes
continue to evoke strong research interest in the general community 
\cite{a1}, since it is potentially a very
powerful probe to investigate the electronic structure of any solid.
Briefly, the Auger process involves the decay of a core-hole by a 
nonradiative transition involving two electrons, one filling the core-hole 
and the second one picking up the excess energy and making a transition 
into the continuum. 
The initial state of the normal Auger transition 
can be considered to be a singly ionized core-hole state and the final state 
is a doubly ionized state together with an electron  in the continuum. 
When a core-hole (C) 
decays via the Auger process involving two valence (V) electrons, 
it is usually termed a C-VV process. 
One can anticipate \cite{1972} the C-VV Auger spectral shape considering
two limiting scenarios. The two final state holes may be found on
different atomic sites due to the presence of finite hopping interactions. 
In such a case they will be screened from each other
and the Auger process can be described 
essentially by
the self-convolution of the occupied
part of the valence band density of states (DOS). In the opposite scenario,
strong correlation effects may localize the holes at the same atomic site,
shifting the corresponding energy higher than the screened two-hole case
by approximately the Coulomb interaction strength, $U$.
These expectations found a theoretical basis
in the work of Cini \cite{cini} and Sawatzky \cite{saw}
who calculated the two-hole Green's function for fully-filled singly degenerate 
Hubbard-like models. The results indeed showed these 
two spectral signatures, often termed the correlated (or the satellite) 
and the uncorrelated (or the main) features. 
According to such analysis, Auger spectroscopy has the unique
advantage of providing a direct experimental estimate of $U$ and such an
approach has often been adopted for the very same purpose \cite{expt}.

The assumptions in the Cini-Sawatzky (CS) theory, however, may limit its
applicability to real systems. The model assumes one orbital per atomic
site, whereas most real systems invariably involve orbital degeneracies.
The other assumption of a fully-filled band limits the use of the model
to few real systems such as Cu and Zn.
It is not {\it a-priori} evident what the consequences of 
relaxing these limitations would be 
on the 
Auger spectra compared to those inferred from the CS theory. Therefore, we
theoretically investigate the Auger spectra from a 
multiband model as a function of  band filling and  Coulomb
interaction strength in order to understand
real systems better and to extend the usefulness of this
very powerful spectroscopic technique. 
We find that the results are {\it qualitatively} 
different for partially filled bands 
compared to the fully-filled case. These strong changes occur primarily
due to the screening response to the core-hole 
in the initial state and multielectron excitations in both initial
and final states.
While these
processes are not possible for a
fully-filled case, our results clearly show that
such processes are in fact the dominant ones in determining the Auger
spectral shapes in the most often encountered partially filled systems.

The specific multiband model that we consider is based on 
a regular tetrahedron with four atomic sites
each containing triply-degenerate $p$-orbitals 
allowing us to treat the full multiplet 
(orbital dependent) electron-electron interactions.
However, the qualitative features presented here are independent
of this specific choice, as we have explicitly checked by calculating the
corresponding spectra with three degenerate $s$ orbitals at each site,
instead of the $p$-orbitals, as well as with five degenerate $d$-orbitals.
The hopping interactions between different orbitals are given in terms of
Slater-Koster parametrization \cite{sk}. 
In order to 
understand the influence of the core-hole and the consequent screening 
dynamics, we have performed two sets of
calculations, one with the core-hole potential and the other
without. The core-hole potential is included in the Hamiltonian in the usual
way \cite{core}, by including the term, -$U_cn_cn_d$, which effectively
lowers the diagonal electronic energy by $U_c$ at the core-hole 
site in the initial
state of the Auger process. 
This assumes that the initial state is a completely screened 
core-hole state. This is reasonable for the transition elements where
the life-time of the core-hole is longer than the screening 
time-scale, but may limit the applicability to other systems,
such as the lanthanides, where these two time-scales may compete
with each other. 
We take  $U_c$ to be 1.2 times the effective
$U$ within the valence band following the usual practice. We present  here
the results for the number ($n$) of electrons per atomic site 
being 1, 2, 3, 4, 5 and 6, with the last one ($n$=6) being the fully-filled 
case. We obtain the ground state and the
corresponding two-hole Green's function by 
the Lanczos and the modified Lanczos methods \cite{lanc}.

In Fig. 1 we show some selected two-hole spectra for 
$n$=6 with a few values of $U/W$. The Auger spectrum for the 
noninteracting limit ($U$=0)
is the self-convolution of the occupied DOS with a width 
equal to twice the occupied part of the single-particle bandwidth, $W$.
With increasing $U$, the intensity within the uncorrelated 
energy region ($\sim$ 0-2.2) rapidly decreases, with increasing 
spectral weight thrown out in a narrow feature.
This narrow feature with its characteristic dependence 
on $U$ has primarily two-holes
at the same site and is easily identified with the
strongly correlated Auger feature of the CS theory.
To underline this similarity further, we plot the average energy
separation (${\Delta}E$) and the intensity ratio ($I_c/I_m$) between the
correlated (satellite) and the uncorrelated (main) 
Auger spectral features as functions of
$U/W$ and $U^2/W^2$, respectively in the 
insets (solid circles) alongwith the
results (open circles) from  the CS
theory. Clearly, both these show  linear 
dependences for larger $U$. Between the two models we find that ${\Delta}E$
are similar even quantitatively. $I_c/I_m$, though qualitatively
similar in both models, is larger in the multiband model.

The spectra from partially-filled bands (Fig. 2) however show
{\it qualitatively} different behavior compared to the 
fully-filled limit. For brevity, we show the results only for some 
selected $n$ and $U/W$ values; we find that the results for other values
of $n$ and $U/W$ are qualitatively similar. 
Figs. 2a and b show the results for
$U/W$=0, 1 and 2 in absence of any core-hole potential ($U_c$=0),
while in Figs. 2c and d, we compare the results with and without
$U_c$ for the $U/W$=1 case. In every case, the spectra appear a lot more
complex than the simple and intuitive expectation of two groups
of features arising from
uncorrelated and correlated final states in the Auger spectrum.
Evidently, the spectra
exhibit multiple groups of distinct and intense spectral features
(marked {\bf 1-5} in the figures for $U/W$=1 case) with as many as four
spectral groups ({\bf 2-5})
appearing  outside the energy region of the 
uncorrelated ($U/W$=0) spectrum (marked {\bf 1}). 
For example, the $U/W$=1 case for $n$=3 (Fig. 2b) has features at about
1.5, 2.8, 4 and 5.3. Such results  cannot be
understood within the CS theory or in terms of the qualitative arguments
presented in the introduction. Additionally, 
${\Delta}E$ and $I_c/I_m$  exhibit no
obvious dependence on $U/W$ and $U^2/W^2$ in 
contrast to the fully-filled case (Fig. 1).
This is most apparent in 
Fig. 2a where the spectral feature {\bf 2} at about 
1.3 moves to a {\it lower} energy with an {\it increase}
of $U/W$ from 1 to 2. 
In order to understand the 
origin of such complex behaviors, we have analysed the character of final-state
two-hole wavefunctions responsible for these various features. We illustrate
schematically 
the dominant contributions to the wavefunctions corresponding to each 
spectral group {\bf 1-5} in Fig. 2d in terms of
electron and hole excitations with respect to the initial state.
The screened, delocalized particles (open and closed circles representing
holes and electrons, respectively) are shown within the semi-elliptical
bandwidth which is arbitrarily chosen to be half-filled for the
purpose of clarity in the presentation. The localized 
hole states are shown in
an atomic-like level.
Thus, the spectral group marked {\bf 1} which corresponds to Auger
transitions appearing within the energy interval of the uncorrelated
($U/W$=0) case corresponds to two screened holes in the band (schematic
{\bf 1} in Fig. 2d).
In each case the large intensity spectral group marked {\bf 2} 
was found to arise from an extra electron-hole excitation accompanying
the generation of two screened holes (see schematics in Fig. 2d). 
The movement of the
feature {\bf 2} to slightly lower energy with increasing $U$ in Fig. 2a is due
to the band-narrowing effects at larger $U$, as the process 
involves an excitation of an electron from the occupied to the
unoccupied part. While the energy position of {\bf 2} is relatively
insensitive to $U$ ({\it e.g.} compare $U/W$=1 and 2 in Fig. 2b),
the existence of this feature is entirely dependent on the presence
of correlation in the initial and final states and consequently,
this feature is completely absent for $U/W$=0. Feature {\bf 3} corresponds
closely to the correlated feature discussed within the CS theory with 
two holes localized at the same atomic site, as shown by the
schematic process {\bf 3} in the Fig. 2d. Thus, this feature moves to
higher energy and rapidly gains intensity with increasing $U$
(compare features {\bf 3} and {\bf 3$^{\prime}$} for $U/W$=1 and 2,
respectively in Fig. 2b). The very weak intensity feature {\bf 4} corresponds to an
electron-hole excitation accompanying the generation of the two 
localized holes, whereas feature {\bf 5} at still higher energy arises
from three holes localized at a site with an electron excited to a
higher energy state. 
Though in these finite size calculations, 
the features {\bf 2} and {\bf 4} appear as distinct peaks, in an extended 
metal such processes will have continuous energy spectra, thus substantially
overlapping the corresponding main peaks, {\bf 1} and {\bf 3} respectively. 
These processes, however, will appear as distinct peaks in the case
of insulators with finite band gaps.
Obviously all these processes ({\bf 2-5} with the 
exception of {\bf 3}) are not describable within any theory for a fully-filled
band, as it depends on multi-particle excitations possible only in
partially-filled bands.

The distinction between the partially-filled and fully-filled bands becomes
all the more striking, when the influence of the core-hole is included. It
is evident that the core hole, existing only in the initial state, does not
have any effect in the fully-filled limit irrespective of the strength of the
core-hole potential ($U_c$), since there cannot be any screening
dynamics here. On the other hand, any reasonable value of 
$U_c$ completely alters the Auger spectrum for any configuration away from
the fully-filled limit, as we show in Figs. 2c and d illustrating typical 
results for $n$=2 and 4 with and
without the core-hole potential.
In both these cases, correlation induced 
features {\bf 2} and {\bf 3} can be seen 
with substantial intensity for $U/W$=1 in the calculated spectra
in absence of a core-hole potential ($U_c$=0; thin solid line)
with weaker features {\bf 4} and {\bf 5} only for $n$= 4 case. For 
$U_c$$\neq$ 0
(thick solid line), however, the intensities in these features all but
vanish completely, transfering almost all the spectral weight to the
uncorrelated spectral range,  {\bf 1}. This observation of disappearance
of the correlation-induced features in presence of $U_c$ is found to be
robust for all reasonable values of $U$ and $U_c$, as explicitly 
checked with extensive calculations. Thus, one obtains here the
paradoxical situation where the pronounced manifestation of
correlation effects within the valence band in 
terms of satellite features ({\bf 2}-{\bf 5})
is virtually wiped out 
{\it due to the presence of another
correlation effect}, namely that between the core hole and the 
valence electrons. In 
order to understand these changes, we note that the 
spectra of the final state energy eigenvalues are identical
for $U_c$=0 and $\neq$ 0 cases, since the core-hole exists
only in the initial state. Therefore, the
drastic modifications in the spectral features 
arise only from changes in the transition probabilities via the 
modification of the initial state wavefunction 
due to the screening response to the 
core-hole potential. 
The effect of the core-hole potential in the initial state 
is primarily to increase the local site occupancy by nearly 1
electron to ($n$+1) as a screening response; analysis of the 
wavefunction character supports this view. Then, 
the subsequent Auger transition creates
2 holes which generate primarily a ($n$-1) configuration
at the local site, and therefore, suppress the correlation induced 
feature arising from  ($n$-2) local
occupancy. 
It is to be noticed that there are some important changes in the 
line-shape of the spectral region {\bf 1} with correlation effects. In absence
of the core-hole, there is a slight narrowing of this spectral region 
arising from band narrowing effects in 
presence of a finite $U$; 
with $U_c \neq 0$, there is a more significant narrowing in this
spectral region (see Fig. 2d). This arises from the above-mentioned changes 
in the transition probabilities transferring weights preferentially to lower 
energies within the uncorrelated region.

In order to ascertain the relevance 
of these results obtained from small, finite 
systems to strongly correlated extended solids, 
we have applied this method of calculation to a
representative case, namely the Auger spectrum of a typical correlated
oxide system, LaCoO$_3$.
The electronic structure of LaCoO$_3$ has been discussed 
extensively in the past literature and there are reliable 
estimates of the various interaction strengths already obtained 
from the analysis of several spectroscopic results \cite{uestim}
within finite cluster many-body calculations involving one
Co and six oxygen atoms in an octahedral geometry. In agreement
with these estimates, we also use the same cluster with the
charge-transfer energy $\Delta$=2.0, $pd{\sigma}$=-1.8,
$A$=4.21, B=0.13, and C=0.64 amounting to a multiplet 
averaged $U_{eff}$ of 4.5 eV.
The calculated Auger spectrum which is the first of its kind 
is compared with experiment \cite{ash}
in Fig. 3, indicating a remarkable agreement. Such an agreement
between the calculated and the experimental spectra without the need 
to readjust parameter values firmly establish the relevance of the
results presented here. More importantly, it leads to a drastic
reinterpretation of the experimental result. It has been
believed that the most intense peak in the Auger spectrum in Fig. 3
at 8.8 eV for LaCoO$_3$ \cite{ash}
arises from Coulomb interaction driven localized two-hole final state,
while the  prominent shoulder at the lower energy 
(5 eV) arises from the screened and delocalized 
two-hole final states. An analysis of the wave-functions 
clearly shows that all the prominent features ({\it i.e.} at  
5 and 8.8 eV) as well as the weak shoulder at about 13.3 eV 
arise from delocalized final states.
We show calculated Auger spectra for LaCoO$_3$ with $U_{eff}$=4.5,
5.5 and 6.5 eV.
Besides minor 
changes, 
the results
are essentially identical inspite of the strong variation in $U$
establishing our conclusions. Additionally, one can see the emergence
of a broad and weak feature beyond 17 eV that moves to higher energy
with increasing $U_{eff}$; only such weak correlation induced 
satellite features survive in the final spectrum. 
In order to understand the origin 
of these spectral features, we first note that the Co $d$ related
single-particle density of 
states is distributed basically in two groups, arising from 
bonding-antibonding splitting of Co $d$ - O $p$ hybridized 
states \cite{band}.
Thus, the self-convolution of the single-particle density of states
has three features corresponding to two holes occupying 
antibonding-antibonding
(A-A), antibonding-bonding (A-B) and bonding-bonding (B-B)
combinations at 4.7, 8.8 and 13.5 eV, establishing the origin
of the experimental features at the same energies to be arising 
from A-A, A-B and B-B occupancies of the delocalized states.
While the energetics
can be explained by this procedure, interestingly the spectral shape cannot 
be described by a self-convolution of the DOS, due to the influence of the 
valence-valence and valence-core interaction effects modifying the transition 
probabilities.
It is found 
that these correlation effects enhance the relative intensity of the peak at
8.8 eV.

In conclusion, we have shown that the Auger spectra from strongly 
correlated, partially-filled systems are profoundly influenced by the
presence of screening dynamics and multielectron processes. 
Satellites in Auger spectra arising from
correlation effects within the valence 
electrons 
is strongly
suppressed by the presence of core-valence correlation effects
leading to the surprising result of 
Auger spectra 
being dominated by
uncorrelated spectral features inspite of the
presence of strong interactions. This calculational method applied for 
the first time to a
real system, LaCoO$_3$ shows a remarkable agreement with the
experimental results, leading to a reinterpretation of Auger 
spectra from such strongly correlated systems.

This research is funded by DST and CSIR, Government of India. We
thankfully acknowledge the use of computational facilities provided by
Prof. S. Ramasesha and Supercomputer Education and Research Centre, Indian
Institute of Science.

\pagebreak
\section{figure captions}

Fig. 1 Calculated Auger spectra for $U/W$=0, 1 and 2 for the fully-filled
case. Insets show the variations in the relative energy position 
(${\Delta}E$) and the
intensity ($I_c/I_m$) of the correlated feature 
compared to the uncorrelated ones
as a function of $U$ and $U^2$, respectively for the multiband
(solid circles) and the single band (open circles) models. All energies 
are in the units of the bandwidth, $W$.

Fig. 2 Calculated Auger spectra for (a) $n$=1, $U/W$=0, 1, 2 and $U_c$=0;
(b) $n$=3, $U/W$=0, 1, 2 and $U_c$=0; (c) $n$=2, $U/W$=0 and $U/W$=1
with and without $U_c$=1.2 $U$; and (d) $n$=4, $U/W$=0 and $U/W$=1 with
and without $U_c$=1.2 $U$. Various groups of spectral features are
marked as {\bf 1-5} and dominant contributions to the corresponding 
wavefunction are shown in panel d schematically. In panel b, high energy
features for $U/W$=1 are multiplied by 10 for clarity. All energies 
are in $W$.

Fig. 3 Comparison of experimental and calculated Co L$_3$-$VV$ Auger spectra
from LaCoO$_3$. The inset illustrates the changes in the calculated
spectra with changing $U$, showing the existence of weak 
and broad correlation induced features at energies above $\sim$ 17 eV.

\begin{references}
\bibitem[*]{} 
Also at the Jawaharlal Nehru Centre for Advanced Scientific Research.
Bangalore. email: sarma@sscu.iisc.ernet.in

\bibitem[\xi]{}
Also at Dept. of Physics, IISc, Bangalore.

\bibitem{auger}
P. Auger, J. Phys. Radium {\bf 6}, 205 (1925).

\bibitem{a1}
D. D. Sarma {\it et al.}, Phys. Rev. Lett. {\bf 63}, 656 (1989);
A. Kivima\"aki {\it et al.}, Phys. Rev. Lett. {\bf 71}, 4307 (1993); 
Z. F. Liu {\it et al.}, Phys. Rev. Lett. {\bf 72}, 621 (1994); 
W. Drube, R. Treusch, and G. Materlik,
Phys. Rev. Lett. {\bf 74}, 42 (1995); S. Aksela, E. Kukk, H. Aksela, 
and S. Svensson, Phys. Rev. Lett. {\bf 74}, 2917 (1995); E. Kukk 
{\it et al.}, Phys. Rev. Lett. {\bf 76}, 3100 (1996).

\bibitem{1972}
S. P. Kowalczyk {\it et al.}, Phys. Rev. B {\bf 8}, 2387 (1973);
E. D. Roberts, P. Weightman, and C. E. Johnson, J. Phys. C {\bf 8}, 2336 (1975). 

\bibitem{cini}
M. Cini, Solid State Commun. {\bf 24}, 681 (1977); Phys. Rev. B {\bf 17},
2788 (1978).

\bibitem{saw}
G. A. Sawatzky, Phys. Rev. Lett. {\bf 39}, 504 (1977).

\bibitem{expt}
Peter A. Bennett {\it et al.}, Phys. Rev. B {\bf 27}, 2194 (1983);
D. K. G. de Boer, C. Haas and G. A. Sawatzky, J. Phys. F {\bf 14}, 2769 (1984);
R. W. Lof {\it et al.}, Phys. Rev. Lett. {\bf 68}, 3924 (1992);
K. Maiti {\it et al.}, Phys. Rev B {\bf 57}, 1572 (1998). 

\bibitem{sk}
J. C. Slater and G. F. Koster, Phys. Rev {\bf 94}, 1498 (1954).

\bibitem{core}
  D.D. Sarma and A. Taraphder, Phys. Rev.  B  {\bf 39},  11570 (1989). 

\bibitem{lanc}
E. R. Gagliano and C. A. Balseiro, Phys. Rev. Lett. {\bf 59}, 2999 (1987);
E. Dagotto, Rev. Mod. Phys. {\bf 66}, 763 (1994).

\bibitem{uestim}
M. Abbate {\it et al.}, Phys. Rev. B {\bf 49}, 7210 (1994);
T. Saitoh {\it et al.}, Phys. Rev. B {\bf 56}, 1290 (1997).

\bibitem{ash}
A. Chainani, M. Mathew and D.D. Sarma, Phys. Rev. B {\bf 46}, 9976 (1992).

\bibitem{band}
D. D. Sarma {\it et al.}, Phys. Rev. Lett. {\bf 75}, 1126 (1995). 

\end{references}
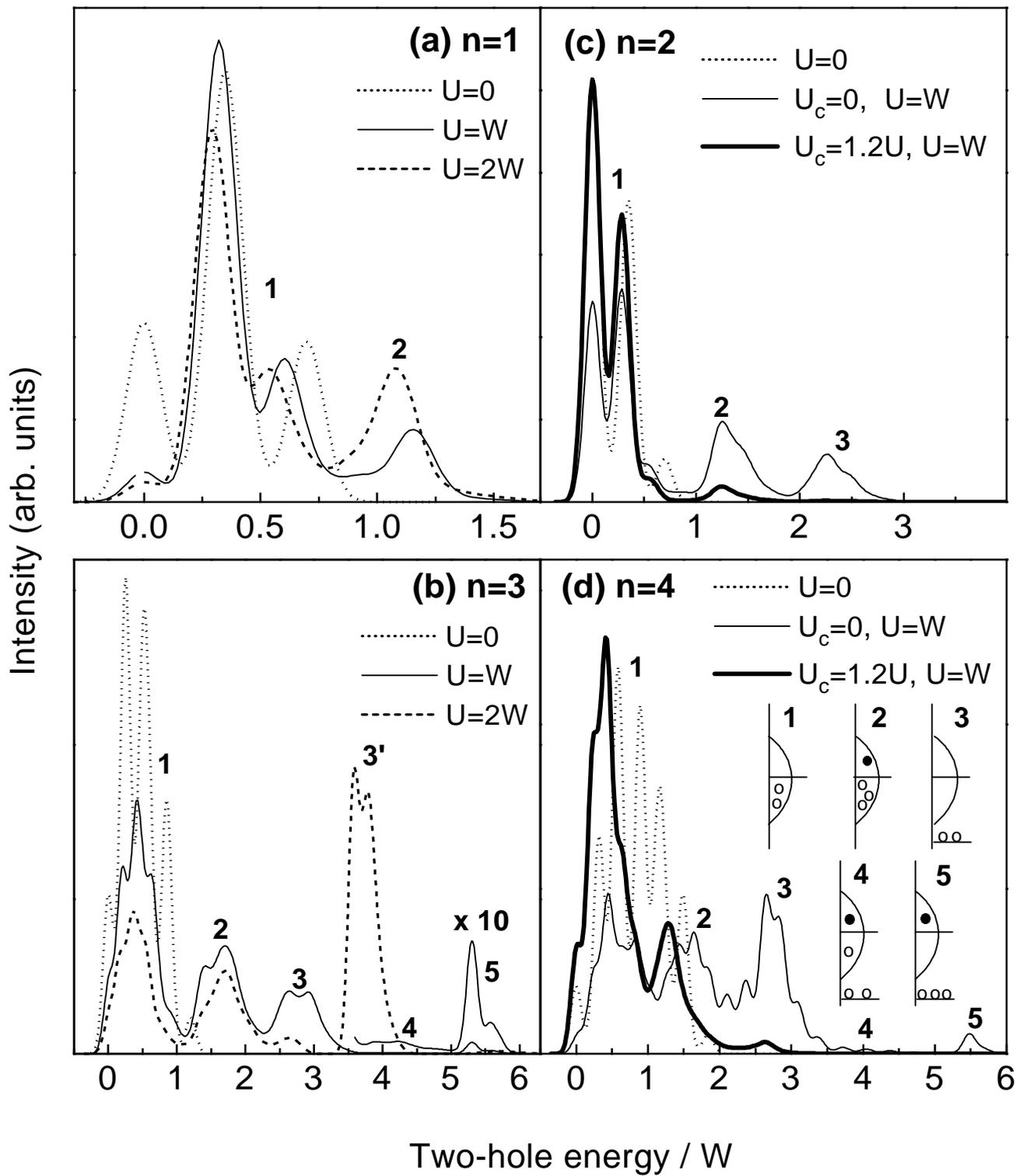
\end{document}